\documentstyle[prl,aps,twocolumn,floats,epsfig]{revtex}
\begin{document}
\draft
\title{
Observation of coupled normal heat and matter
transport in a simple model system
}

\author{C. Mej\'\i a--Monasterio, H. Larralde and F. Leyvraz}
\address{Centro de Ciencias F\'\i sicas, av.~Universidad
s/n, Colonia Chamilpa, Universidad Nacional Aut\'onoma de Mexico,
Cuernavaca, Morelos, Mexico}
\date{\today}
\maketitle
\begin{abstract}
We introduce the first simple mechanical system that shows fully
realistic transport behavior while still being exactly solvable at the
level of equilibrium statistical mechanics.  The system under
consideration is a Lorentz gas with fixed freely-rotating circular
scatterers which scatter point particles via perfectly rough
collisions. Upon imposing either a temperature gradient and/or a
chemical potential gradient, a stationary state is attained for which
local thermal equilibrium holds. Transport in this system is normal,
in the sense that the transport coefficients which characterize the
flow of heat and matter are finite in the thermodynamic
limit. Moreover, the two flows are non-trivially coupled, satisfying
Onsager's reciprocity relations to within numerical accuracy.
\end{abstract}
\pacs{05.60.-k, 05.20.-y, 44.10.+i}

The phenomenological equations of irreversible thermodynamics, which
describes diffusion, heat conduction, viscosity among a host of other
phenomena, are of great significance in many fields. To link these
phenomena to the underlying microscopic dynamics is, however, highly
non trivial and the results so far have not been conclusive. From a
mathematically rigorous point of view, very few results have been
obtained \cite{bonetto}. From the numerical point of view, the systems
considered thus far are either too complicated to shed much light upon
the problem, or fail to reproduce the macroscopic phenomenology.
Indeed, to our knowledge, the validity of Fourier's law has been
proven analytically only for a specific model in the limit of infinite
dilution with finite mean free path \cite{lebowitz1,lebowitz2}.  There
have also been attempts to link transport phenomena to the chaotic
properties of the underlying classical dynamics
\cite{chaos}, and a connection between the
rate of entropy production and the rate of contraction of phase space
volume has also been pointed out \cite{chernov}.  Further results have
also been obtained concerning energy transport in the Lorentz gas
\cite{alonso}. However, as was shown in \cite{dhar}, this last system
does not satisfy local thermal equilibrium (LTE), which is a
fundamental assumption of irreversible thermodynamics. It is therefore
not clear what precise meaning should be attached to Fourier's law in
this situation.  Numerically, attempts have been made, on one side,
through the simulation of realistic many-body systems satisfying a
thermostatted (non Hamiltonian) dynamics \cite{evans}.  These
simulations have indeed been able to reproduce non-trivial transport
phenomena.  However, such studies do not provide a detailed
understanding of the microscopic processes involved due to the
excessive complexity of the system.  The other numerical approach
involves the study of transport in ``simple systems''. Examples of
these include: chains of anharmonic oscillators
\cite{lepri}, for which Fourier's law does not hold; Lorentz gases
\cite{alonso}, which, as mentioned above, do not satisfy LTE; and the
so called ding-a-ling and ding-dong models \cite{casati,prosen1},
which indeed yield normal heat transport, but have geometric
constraints that make even their equilibrium properties a complicated
affair and cannot be generalized, retaining their simplicity, to
obtain the coupled heat and matter transport commonly found in real
systems. Under these circumstances, the minimal ingredients required
in the microscopic physics of a system to attain normal heat
conduction in low dimensions are presently still under discussion
\cite{bonetto,hu,prosen2}.

In this work we introduce the first simple reversible Hamiltonian
model system whose steady state is well described by the hypothesis of
LTE and supports both normal heat and matter transport. These are
non-trivially coupled, satisfying Onsager's reciprocity
relations. Furthermore, this system has the advantage that its
equilibrium properties are trivial since its description reduces to
that of ideal gases. This makes our model an ideal testing ground for
theories linking microscopic mechanical properties to transport
phenomena. Moreover, this model can be easily modified to study other
physical problems from a microscopic approach, such as transport
in heterogeneous systems \cite{mahan} and the effects of applied fields.

Our model consists of non-interacting point particles of mass $m$
scattered by an array of {\it freely-rotating\/} circular scatterers
of radius $R$ with a finite moment of inertia $\Theta$.  The only
relevant adimensional parameter characterizing the system is
\begin{equation}
\eta=\frac{\Theta}{mR^2} \ ,
\label{eq:defeta}
\end{equation}
which determines the amount of energy transfer between disks and
particles in a collision.  The scattering proceeds according to the
rules characterizing perfectly rough collisions, designed to
conserve energy and angular momentum. They are given by the following
prescription: the normal velocity $v_n$ of the particle changes sign, whereas
the tangential velocity $v_t$ of the particle and the angular velocity
$\omega$ of the disk are transformed as follows:

\begin{eqnarray}
v_n' & = & -v_n\, \nonumber\\
v_t' & = &  v_t - \frac{2\eta}{1 + \eta} (v_t - R\omega)\,  \\
\label{eq.collision}
R\omega'& = &  R\omega + \frac{2}{1 + \eta} (v_t - R\omega) \nonumber \ .
\end{eqnarray}

These rules define a deterministic, time-reversible, and canonical
transformation at each collision.  For finite values of $\eta$,
particles in the system may exchange energy among each other through
the disks, even though they do not interact directly.  It is precisely
this disk-mediated energy exchange which permits the systems to reach
thermodynamical equilibrium without the necessity of any additional
thermostat (a similar idea was proposed in \cite{klages} as a model of
a deterministic thermostat). It should be noted that both the limits
$\eta\to0$ and $\eta\to\infty$ are exceptional: In the former case the
particle dynamics reduces to that of normal elastic scattering by a
hard disk, which is the usual Lorentz gas model; in the latter, the
state of rotation of the discs is unaffected by the collisions. Thus,
in both limits the energy-mediating effect of the disks is suppressed
and thermodynamical equilibrium is not reached. In the following,
unless the contrary is explicitly stated, we shall always be dealing
with the case $\eta=1$, since that is the value of $\eta$ for which
equilibration is most efficient. We shall also always set $m$ and $R$
to one.

A peculiarity of this system is that it is homogeneous, i.e.  it has
no proper energy scale, or, equivalently, no proper time scale.  Thus
all energies and temperatures reported in this work are rescaled to
the lower nominal temperature of the baths.

\begin{figure}[!t]
\centerline{\epsfxsize=0.95\columnwidth \epsffile{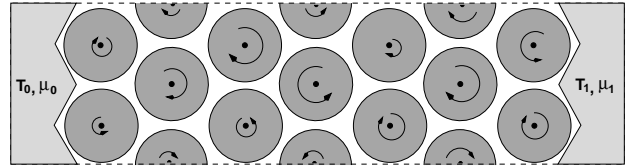}}
\vspace{0.5cm}
\caption{Schematic illustration of the scatterer geometry: the scatterers are
disposed on a triangular array with finite horizon to avoid infinitely
long trajectories. For matters of convenience, in this work the
separation between scatterers has been set to have the critical
horizon. Periodic boundary conditions are used in the vertical
direction. To avoid spurious effects arising from multiple scattering
off the same disk, we have put two disks on each vertical. To study
the dependence with system size, the length L of the sample is
varied. The quoted length is the number of disks.}
\label{fig.1}
\end{figure}

The geometric disposition of the scatterers is indicated in
Fig.~\ref{fig.1}.  The centers of the scatterers are fixed on a
triangular lattice, along a narrow channel with periodic boundary
conditions in the vertical direction.  At the ends, the walls are used
to fix both the temperature $T$ and the quantity $\mu/T$, where $\mu$
is the chemical potential. This is achieved by ensuring that each
particle that hits the right or the left wall is absorbed with
probability $1-\exp({-1/|v_n|})$, where $v_n$ is the normal
component to the wall of the particle velocity. Otherwise, the
particle is reflected with a velocity chosen from
\begin{eqnarray}
P_n(v_n) & = & \frac{1}{T}|v_n|\exp\left(-\frac{v_n^2}{2T}\right) \ , \nonumber\\
P_t(v_t) & = & \frac{1}{\sqrt{2\pi
T}}\exp\left(-\frac{v_t^2}{2T}\right) \ ;
\label{dist}
\end{eqnarray}
where $v_n$ and $v_t$ are, respectively, the normal and tangential
components of the velocity with respect to the wall, and $T$ is the
nominal temperature of the wall. Additionally, the walls emit particles
with velocities distributed as in (\ref{dist}), at a velocity dependent
rate given by $\gamma(1-\exp({-1/|v_n|})$. Here, $\gamma$ determines
the nominal value of the chemical potential at the walls through
$\mu=T\ln({\gamma/T^{3/2}})$. Details will be given in a forthcoming
publication \cite{mejia}.
At the level of equilibrium statistical mechanics the system is an
ideal gas mixture, with the disks having one degree of freedom and the
particles two.  We have verified that our model reaches an equilibrium
state fulfilling the equipartition theorem in the microcanonical,
canonical and grandcanonical ensembles of equilibrium statistical
mechanics. Further, the temperature and chemical potential at which
the system equilibrates coincides with the nominal values 
at the walls in the canonical and grand
canonical cases. Finally, when the system is subjected to a weak
temperature or chemical potential gradient, it reaches a well-defined
stationary state after a relaxation time that depends on the length of
the system and on the gradients themselves.
\begin{figure}[!b]
\centerline{\epsfxsize=0.95\columnwidth \epsffile{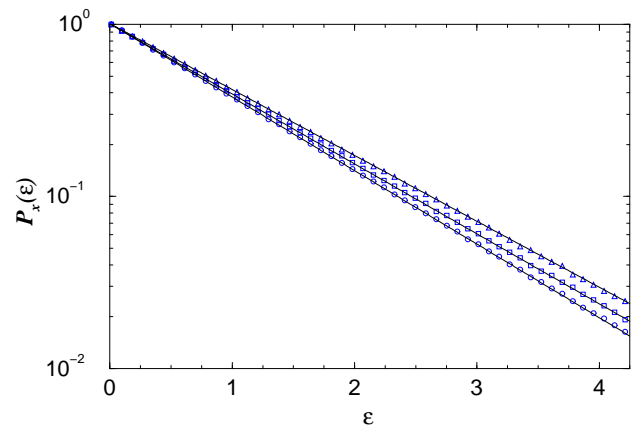}}
\caption{Semilogarithmic plot of the particle's energy distribution
$P_x(\varepsilon)$ at different positions along a channel of length $30$
with an end-to-end temperature difference $\Delta T=1/7$ and
$\Delta(\mu/T)=-0.2002$. The different curves correspond to a fit to a
Boltzmann distribution for each position. From these fits, we obtain
the temperatures: $T = 1.0193$ at $x = 3$ (circles), $T = 1.0693$ at $x
= 13$ (squares) and $T = 1.1335$ at $x = 27$ (triangles). The curves
have been rescaled for clarity.}
\label{fig.2}
\end{figure}
To show that LTE is indeed achieved in the stationary state attained
by our system, we display in Fig.~\ref{fig.2} a typical realization of
the energy distribution function $P_x(\varepsilon)$ of the particles
as they cross a narrow slab centered around position $x$ for three
different positions. At each position, $P_x(\varepsilon)$ is
consistent with the Boltzmann distribution.  Further, if we determine
the temperature $T(x)$ by a fit of the $P_x(\varepsilon)$ to the
Boltzmann distribution, the agreement with the expected linear
temperature profile in the system is satisfactory and coincides with
the average energy of the particles measured at that position, see the
inset of Fig.~\ref{fig.3}. Thus, the identification of the average
particle energy with the local temperature is justified in this
system.
\begin{figure}[!t]
\centerline{\epsfxsize=0.95\columnwidth \epsffile{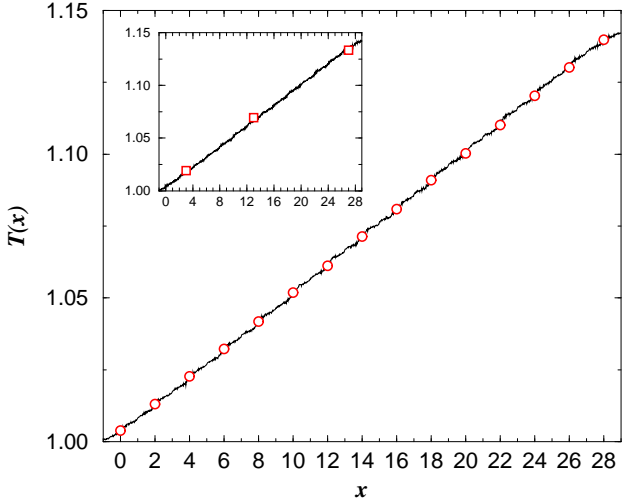}}
\caption{Temperature profile for the simulation described in
Fig.~\ref{fig.2}. The solid line corresponds to the particle
temperature along the channel while the open circles are the
temperature of the disks, averaged over 500 realizations.  In the
inset the particle temperature profile (solid line) is compared with
the temperatures obtained from the fits to the Boltzmann distribution
of $P_x(\varepsilon)$ in Fig.~\ref{fig.2}, (open squares).}
\label{fig.3}
\end{figure}
In Fig.~\ref{fig.3} we show the temperature profiles of both particles
and disks, which are found to be linear. The agreement between both
values indicates that disks equilibrate with the particles locally
along the channel.  In the inset, the temperatures obtained in
Fig.~\ref{fig.2} for three different positions are compared with
$T(x)$, measured as the average energy. The agreement reinforces the
conclusion that the system establishes a LTE in its steady state. Note
again that $T(x)$ reaches the nominal temperatures specified by the
walls at both ends. This is in contrast to what is observed in several
other models for heat conduction previously proposed
\cite{prosen1,lepri,hu}.  Moreover, in simulations in which we impose
a temperature gradient but the same value of $\mu/T$ on both walls, we
find that $\rho(x)/T(x)$ is constant along the channel, confirming
that the particles in the system can be described as a
two-dimensional ideal gas which is locally at equilibrium.

From the general theory of irreversible processes, to linear order
the heat and particle currents $J_u$ and $J_\rho$ can be written as
follows (see e.g. \cite{reichl}):
\begin{eqnarray}
J_u&=&L_{uu}\nabla\frac{1}{T}-L_{u\rho}\nabla\frac{\mu}{T}\nonumber \ ,\\
J_\rho&=&L_{\rho u}\nabla\frac{1}{T}-L_{\rho\rho}\nabla\frac{\mu}{T}
\label{eq:const}
\end{eqnarray}
and the Onsager reciprocity relations read in this case
\begin{equation}
L_{u\rho}=L_{\rho u} \ .
\label{eq:recip}
\end{equation}

\begin{figure}[!t]
\centerline{\epsfxsize=0.95\columnwidth \epsffile{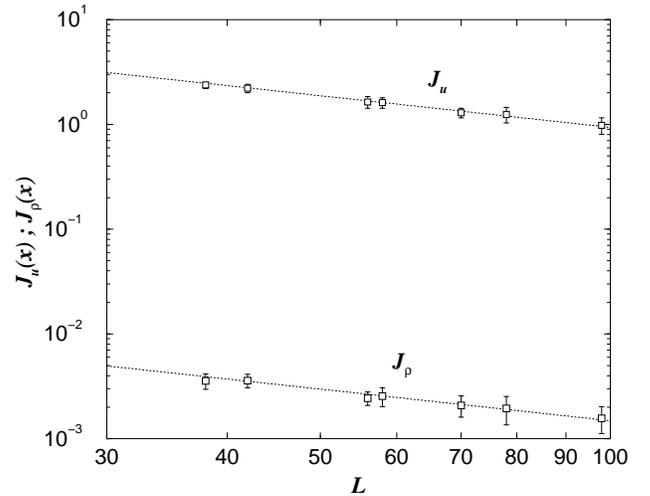}}
\caption{Size dependence of the heat and matter currents for
simulations with a fixed temperature difference, $\Delta T=1/7$, and
$\mu/T$ constant. The dotted lines corresponds to $1/L$ scaling.  From
the agreement it follows that the transport coefficients do not depend
on the size of the system.}
\label{fig.5}
\end{figure}

The central feature of the model is that its transport properties are
{\it normal}, meaning that the various transport coefficients
appearing in (\ref{eq:const}) do not depend on the length of system,
and are, thus, well defined in the thermodynamical limit. In
Fig.~\ref{fig.5} we show the dependence of the currents on the length
$L$ of the system, for a typical realization. The $1/L$ dependence
observed confirms that transport is normal.

In order to obtain the value of the coefficients in
(\ref{eq:const}), it is enough to perform two simulations: Fixing
the value of $\nabla T$ and setting $\nabla(\mu/T)=0$ yields $L_{uu}$
and $L_{\rho u}$ while setting $\nabla T =0$ and fixing
$\nabla(\mu/T)$ gives $L_{u\rho}$ and $L_{\rho \rho}$.

We have performed simulations with temperature and chemical potential
differences up to 20\% of the minimal nominal values at the walls, and
in all cases we have found normal transport consistent with
(\ref{eq:recip}).
For example, in a simulation in which a temperature difference of
$\Delta T=1/14$ and $L=30$ at fixed $\mu/T=-5.01$, averaged over 500
realizations, we found that after the steady state has been reached,
both a heat current and a particle current were driven by the
temperature gradient.  The measured value for $L_{\rho u}$ in this
case was $0.0054\pm 0.00013$. This implies, through (\ref{eq:recip}),
that a heat current will be driven by a gradient in $\mu/T$ at fixed
$T$, as is indeed observed.  In the complementary simulation with $T$
constant, where $\Delta\frac{\mu}{T}=-0.06$, we found $L_{u
\rho}=0.0055\pm0.00013$, thus confirming (\ref{eq:recip}) to within
our numerical accuracy.

As a consistency check, we have also studied a ``canonical''
situation, in which we supressed absorption and emission of particles
at the walls. In this situation there is no flow of matter in the
steady state. The relationship between heat flow and temperature
gradient becomes $J_u=\kappa \nabla T$, with $\kappa$ given by the
following expression:
\begin{equation}
\kappa=\frac{L_{uu}L_{\rho\rho}-L_{u\rho}L_{\rho u}}{T^2L_{\rho\rho}} \ .
\label{eq:canonical}
\end{equation}
This relationship was found to hold to good accuracy, thus confirming
the validity of our ``grand canonical'' simulations by which the $L$'s
were evaluated.

We argue that the coupling between the two currents is non-trivial in
the following sense: in a ``canonical'' simulation, the simplest
assumption for the dependence of the density on the position is that
the orbit of each particle covers the sample uniformly. Then the local
temperature merely determines the speed at which the orbit is being
traversed. This would imply that in such a simulation, the particle
density scales inversely with the average velocity, that is
\begin{equation}
\rho(x)T(x)^{1/2}=const \ .
\label{eq:inconstancy}
\end{equation}

In terms of the transport coefficients defined in (\ref{eq:const}),
equation (\ref{eq:inconstancy}) is equivalent to $L_{\rho
u}=(d+1)TL_{\rho\rho}/2$, where $d$ is the dimension of the system.
This relation corresponds to a system for which all transport arises
from uncorrelated Markovian motion of the particles, as in the Knudsen
gas \cite{reichl}.

However, equation~(\ref{eq:inconstancy}) does not hold in our system as we
always find a systematic spatial variation in this quantity, the size
and sign of which depend on the value of $\eta$. Thus, our system
cannot be accurately described in this simple manner. The correct
description is, as in all systems showing realistic transport
properties, an open problem.

In summary, we have introduced a reversible Hamiltonian system to
study transport, with the following desirable properties: Its
statistical mechanics is trivial, reducing to that of ideal gases. In
the steady state it is well described by the hypothesis of local
thermal equilibrium. Its transport properties are normal and it
naturally supports coupled matter and heat transport for which
Onsager's reciprocity relations are satisfied to within numerical
accuracy. This coupling is non-trivial in that it does not reduce to a
simple velocity scaling. In view of these properties, we believe that
this model provides an ideal framework from which theories for the
emergence of macroscopic transport in out-of-equilibrium conditions
can be constructed and/or tested. Furthermore, the model opens the
possibility to study other phenomena such as thermal junctions (and
other kinds of inhomogeneities) by changing the size or moment of
inertia of the disks, Joule heating by applying an electric field,
broken time reversal symmetry by applying magnetic fields, etc
\cite{mazur}. Research into some of these extensions is currently
underway \cite{mejia}.

We acknowledge useful discussions with G. Casati, S. Ruffo, T. Prosen
and T. H. Seligman.  We are thankful to S. Redner, L. Moch\'an for
careful revisions to the manuscript.  H.L. and F.L. acknowledge
partial financial support by DGAPA-UNAM.  C.M.  acknowledges
a fellowship by DGEP-UNAM.

\end{document}